\documentclass[10pt,conference]{IEEEtran}
\IEEEoverridecommandlockouts
\usepackage{cite}
\usepackage{amsmath,amssymb,amsfonts}
\usepackage{algorithm}
\usepackage{algorithmic}
\usepackage{graphicx}
\usepackage{textcomp}
\usepackage{arydshln}
\usepackage{url}
\usepackage{hyperref}
\usepackage{xcolor}
\def\BibTeX{{\rm B\kern-.05em{\sc i\kern-.025em b}\kern-.08em
    T\kern-.1667em\lower.7ex\hbox{E}\kern-.125emX}}

\usepackage{xcolor}
\usepackage{arydshln}
\usepackage {booktabs}
\definecolor{EMgray}{gray}{0.45}

\begin{document}

\title{Rethinking Masking Strategies for Masked Prediction-based Audio Self-supervised Learning
\thanks{This work was partially supported by JST Strategic International Collaborative Research Program (SICORP), Grant Number JPMJSC2306, Japan.}
}

\author{\IEEEauthorblockN{Daisuke Niizumi$^\dagger$$^\ddagger$$^\P$, Daiki Takeuchi$^\ddagger$, Masahiro Yasuda$^\ddagger$, Binh Thien Nguyen$^\ddagger$,\\ {Noboru Harada$^\ddagger$, and Nobutaka Ono$^\dagger$}}
\IEEEauthorblockA{$^\dagger$\textit{Tokyo Metropolitan University}, Tokyo, Japan\\
\IEEEauthorblockA{$^\ddagger$\textit{NTT Inc.}, Atsugi, Japan}
$^\P$daisukelab.cs@gmail.com}
}

\maketitle

\begin{abstract}
Since the introduction of Masked Autoencoders, various improvements to masking techniques have been explored. In this paper, we rethink masking strategies for audio representation learning using masked prediction-based self-supervised learning (SSL) on general audio spectrograms. While recent informed masking techniques have attracted attention, we observe that they incur substantial computational overhead. Motivated by this observation, we propose dispersion-weighted masking (DWM), a lightweight masking strategy that leverages the spectral sparsity inherent in the frequency structure of audio content. Our experiments show that inverse block masking, commonly used in recent SSL frameworks, improves audio event understanding performance while introducing a trade-off in generalization. The proposed DWM alleviates these limitations and computational complexity, leading to consistent performance improvements. This work provides practical guidance on masking strategy design for masked prediction-based audio representation learning.
\end{abstract}


\section{Introduction}
Masked Autoencoders (MAE)\cite{he2022masked} have had a significant impact on representation learning through masked prediction-based self-supervised learning (SSL), initially in the image domain, and have subsequently influenced various studies on general-purpose audio representations\cite{Baade2022MAE-AST,niizumi2022msm-mae,huang2022amae}.
While the original MAE demonstrated the effectiveness of fully random masking, subsequent work in the image domain showed that more structured strategies can lead to improved representations, such as block masking in BEiT\cite{bao2021beit} and inverse block masking (IBM) in data2vec 2.0\cite{Baevski2023data2vec2}.
In addition to masking strategies that rely purely on randomness regardless of data content, several recent studies have explored informed masking approaches\cite{Li2022SemMAE,Jeongwoo2024SelfGuided}, which exploit attention patterns or patch-level representations derived from pretrained models or from the model itself during training.

In the audio domain, many existing methods largely follow the random masking strategy introduced in MAE, whereas recent SSL approaches have reported performance gains by adopting IBM\cite{EAT_SSL,AaPE2025}.
However, in these studies, the contributions of masking strategies are often entangled with other training components, and evaluations are typically conducted on a limited set of downstream tasks.
As a result, the role of masking strategy design in masked prediction-based audio representation learning remains insufficiently understood, especially in terms of generalization performance, leaving room for further investigation into alternative masking strategies.

\begin{figure}[tbp]
  \centering
  \includegraphics[width=1.0\columnwidth]{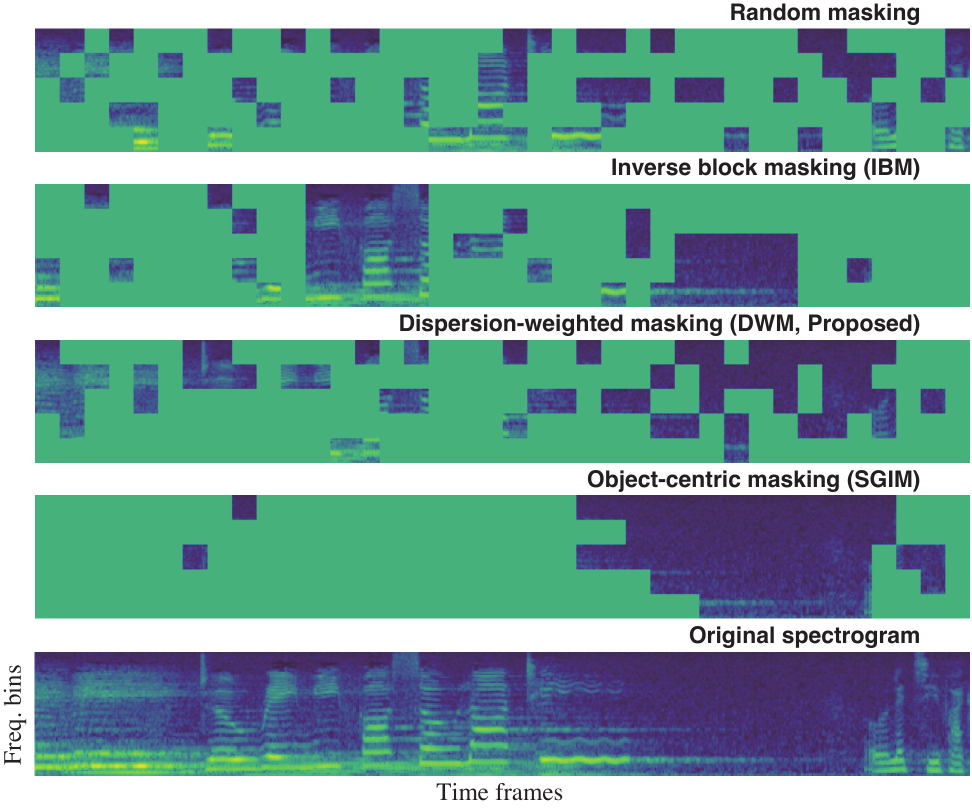}
  \vspace{-10pt}
  \caption{Masking examples on an audio spectrogram (hint ratio $r_h=0.0$). Object-centric masking (SGIM) almost entirely covers audio events (objects), whereas the proposed DWM applies dispersion-weighted random masking based on patch-wise spectral variability, yielding masking patterns that loosely align with object-centric behavior.}
  \label{fig:mask_samples}
  \vspace{-10pt}
\end{figure}

In this study, we investigate how various masking strategies, ranging from random masking to informed masking, contribute to masked prediction-based representation learning for general audio (e.g., environmental sound, speech, and music) using spectrogram inputs.
Using three masked prediction-based SSL frameworks, we conduct a comparative analysis of random masking, IBM, informed masking, and the proposed dispersion-weighted masking (DWM), as illustrated in Fig.~\ref{fig:mask_samples}.

DWM is motivated by the observation that typical audio events do not occupy all frequency components uniformly, resulting in blank or low-energy regions in spectrogram frequency bins.
By leveraging this property, DWM assigns higher masking probabilities to regions with larger dispersion, thereby incorporating spectral structure into the masking process.
Compared to previous informed masking approaches that require non-negligible computational overhead to identify candidate regions, DWM provides a simple and lightweight algorithm, enabling a spectrogram-friendly and efficient form of informed masking.

Our experimental results provide insights into how different masking strategies affect representation learning.
While IBM facilitates the learning of representations that are well suited for audio event semantics, it degrades performance on other downstream tasks, resulting in an overall trade-off.
Informed masking approaches, on the other hand, incur non-negligible computational costs due to the need to compute masking candidates using the model during training.
In contrast, the proposed DWM improves performance on audio event understanding without increasing computational complexity, while suppressing performance degradation on other tasks and preserving generalization performance.

The findings of this study, together with the proposed DWM, are applicable to a broad range of masked prediction-based audio representation learning methods and contribute to masking strategy design in the following ways:

\begin{itemize}
\item We clarify the impact of masking strategies on audio spectrogram-based representation learning, highlighting their effects across different downstream tasks.
\item We propose DWM, a lightweight masking strategy tailored to audio spectrograms, which leverages dispersion patterns with minimal additional computational cost.
\end{itemize}

We publicly release our code\footnote{\url{https://github.com/onolab-tmu/audio_ssl_masking}} to support reproducibility and to facilitate further advances in masked prediction-based audio representation learning.

\section{Preliminary}
We introduce the background and prior work necessary to frame our study on masking strategies for masked prediction-based representation learning in this section.

\subsection{Masked prediction-based representation learning for audio spectrograms}
Masked prediction-based representation learning operates by dividing an input into visible and masked patches, denoted as $x_v$ and $x_m$, respectively.
An encoder $f(\cdot)$ processes only the visible patches to produce latent representations $z_v = f(x_v)$.
A predictor $g(\cdot)$ then infers the masked part of the input as
$\hat{z}_m = g(\mathrm{concat}(z_v, m))$,
where $m$ denotes mask tokens concatenated at the masked positions.

The form of the prediction target $\hat{z}_m$ and the role of the predictor $g(\cdot)$ depend on the specific method.
For example, in MAE\cite{he2022masked}, $\hat{z}_m$ corresponds to the reconstructed input signal in the masked regions, and $g(\cdot)$ acts as a reconstruction decoder.
In Masked Modeling Duo (M2D)\cite{M2D2024TASLP}, also used in our experiments, $\hat{z}_m$ represents latent representations of the masked regions, and $g(\cdot)$ serves as a predictor in the representation space.

For audio spectrograms, MAE-AST\cite{Baade2022MAE-AST}, MSM-MAE\cite{niizumi2022msm-mae}, Audio-MAE\cite{huang2022amae}, and CED\cite{dinkel2023ced} are MAE-based methods.
Prior to MAE, SSAST\cite{gong2022ssast} also learns representations using a reconstruction-based objective.
In contrast, M2D\cite{M2D2024TASLP}, ATST\cite{Li2023ATST-TALSP}, EAT\cite{EAT_SSL}, and AaPE\cite{AaPE2025} learn by predicting masked patch representations, while BEATs\cite{chen2022beats} predicts tokenized labels. M2D-CLAP\cite{niizumi2025m2d-clap} further extends M2D by combining masked prediction with contrastive language-audio pre-training (CLAP), enabling the model to directly learn semantic information from natural language descriptions of audio.
Among these methods, most adopt a random masking strategy.
More recently, EAT and AaPE have demonstrated performance improvements by leveraging IBM.

\subsection{Masking strategies}

Among existing masking strategies, we consider random masking, inverse block masking (IBM)\cite{Baevski2023data2vec2}, and self-guided informed masking (SGIM)\cite{Jeongwoo2024SelfGuided} in our evaluation.

\paragraph{Random Masking}
In MAE, several masking variants, such as block and grid masking, were investigated, and random masking was found to be the most effective, enabling higher masking ratios while maintaining strong performance.

\paragraph{Inverse Block Masking}
IBM starts with a complete mask over all patches and iteratively restores original content in block-shaped regions, which may overlap, until the number of masked embeddings matches a target masking ratio.
As a result, visible patches remain in contiguous block regions; however, the selection of these regions is performed randomly and independently of the input content, making IBM a content-agnostic masking strategy.
EAT\cite{EAT_SSL} extends IBM to audio spectrograms and provides empirical evidence of its utility for audio representation learning.

\paragraph{Self-Guided Informed Masking}
SGIM, proposed in SG-MAE\cite{Jeongwoo2024SelfGuided}, is an object-centric masking strategy that applies masks to estimated object-centric regions. 

SGIM assigns an object relevance score $S_i$ to each of the $N$ input patches, where $i$ denotes the patch index, and determines masked and visible patches based on their ranking.
A hint ratio is used to randomly replace a subset of masked patches with visible ones to control task difficulty, and is gradually annealed toward zero through scheduling, as in \cite{Li2022SemMAE}.

To compute these relevance scores, SGIM performs a Normalized Cut\cite{NCut}-based partitioning using pairwise similarities between patch features\cite{SSL_NCut}.
Specifically, it applies an eigenvalue decomposition to the similarity matrix constructed from patch representations, and the resulting partition is used to evaluate the relevance score of each patch.

However, this procedure involves eigenvalue decomposition, whose computational complexity generally scales as $O(N^3)$.
This high computational cost poses a challenge to training efficiency in self-supervised pre-training, where a large number of patches and batch-wise processing are required.

In our preliminary implementation based on M2D using four A100 GPUs, random masking required approximately 7 minutes per epoch, whereas SGIM incurred a substantially higher cost of approximately 35 minutes, resulting in roughly a fivefold slowdown.
Due to this computational overhead, applying SGIM broadly across masked prediction-based learning frameworks can be impractical.

These limitations motivate the need for a lightweight alternative that retains the benefits of informed masking while remaining computationally efficient.

\section{Proposed Masking Method: DWM}
DWM aims to approximate informed masking tailored to audio spectrograms while maintaining low computational cost.
As illustrated by the object-centric masking example produced by SGIM in Fig.~\ref{fig:mask_samples} (Object-centric masking), patches corresponding to spectrally sparse regions are often assigned as visible in SGIM.
Unlike images, audio spectrograms naturally contain blank regions when certain frequency components are absent.
Based on this observation, DWM is motivated by the idea that the dispersion of a spectrogram patch can serve as a proxy for the amount of information related to audio events contained in that patch.

As shown in Algorithm~\ref{alg:DWM}, DWM operates on an input patch sequence $X$ and first estimates patch-wise importance using mean absolute deviation (MAD) as a dispersion metric.
The MAD of a patch $x_i$ is defined as $\text{MAD}(x_i) = \frac{1}{n}\sum_{j}|x_{i,j} - \bar{x}_i|$, where $n$ is the number of elements in the patch and $\bar{x}_i$ is the patch mean, and $\epsilon$ is a small constant to prevent division by zero.
The resulting dispersion values are converted into sampling probabilities, which are used to sample an initial set of masked patches $M_0$ in a weighted manner.
From $M_0$, a subset of patches is randomly selected according to the hint ratio and returned to the visible set.
To preserve the target mask ratio, the same number of patches is then randomly selected from the visible set and reassigned as masked.

As a result, patches with higher dispersion are probabilistically more likely to be masked, while the hint-based exchange process prevents the masking task from becoming overly difficult.
Although DWM does not explicitly model object-centric regions, it assigns lower masking probability to spectrally sparse patches, which partially aligns its masking behavior with the effects observed in SGIM for audio spectrograms, without expensive computations.

Similar to SGIM, the hint ratio $r_h$ is determined by a scheduling strategy. Following SemMAE\cite{Li2022SemMAE}, $r_h$ is updated based on training progress as 
\begin{equation}
r_h = 1.0 - \left(\frac{\text{epoch}}{\text{total\_epochs}} \right)^{\gamma},
\end{equation}
where $\gamma$ is a scheduling parameter.

\begin{algorithm}[t]
\caption{Dispersion-weighted Masking (DWM)}
\label{alg:DWM}
\begin{algorithmic}[1]
\REQUIRE Input sequence $X$, mask ratio $r_m$, hint ratio $r_h$.
\ENSURE Visible indices $\mathcal{V}_{\text{final}}$, Masked indices $\mathcal{M}_{\text{final}}$.

\STATE \textit{\color{EMgray}\# Step 1: Setup parameters}
\STATE $L \gets \text{Total number of patches}$
\STATE $N_{\text{keep}} \gets \lfloor L \cdot (1 - r_m) \rfloor$, $N_{\text{mask}} \gets L - N_{\text{keep}}$
\STATE $N_{\text{hint}} \gets \lfloor N_{\text{mask}} \cdot r_h \rfloor$

\STATE \textit{\color{EMgray}\# Step 2: Importance estimation}
\FOR{each patch $x_i$ in $X$}
    \STATE $\omega_i \gets \text{MAD}(x_i)$ \COMMENT{Calculate mean absolute deviation}
\ENDFOR
\STATE $P(i) \gets (\omega_i + \epsilon) / \sum (\omega_j + \epsilon)$ \COMMENT{Sampling probability}

\STATE \textit{\color{EMgray}\# Step 3: Initial weighted sampling}
\STATE $\mathcal{M}_0 \gets \text{Sample}(\{1, \dots, L\}, N_{\text{mask}}, \text{prob}=P)$
\STATE $\mathcal{V}_0 \gets \{1, \dots, L\} \setminus \mathcal{M}_0$

\STATE \textit{\color{EMgray}\# Step 4: Hint-based exchange process}
\STATE $\mathcal{H} \gets \text{RandomSelect}(\mathcal{M}_0, N_{\text{hint}})$ \COMMENT{Select hints}
\STATE $\mathcal{V}_{\text{temp}} \gets \mathcal{V}_0 \cup \mathcal{H}$
\STATE $\mathcal{C} \gets \text{RandomSelect}(\mathcal{V}_{\text{temp}}, N_{\text{hint}})$ \COMMENT{Maintain $N_{\text{keep}}$}

\STATE \textit{\color{EMgray}\# Step 5: Final index sets}
\STATE $\mathcal{V}_{\text{final}} \gets \mathcal{V}_{\text{temp}} \setminus \mathcal{C}$
\STATE $\mathcal{M}_{\text{final}} \gets \{1, \dots, L\} \setminus \mathcal{V}_{\text{final}}$
\end{algorithmic}
\end{algorithm}

\section{Experiments}
We investigated the effects of different masking strategies on masked prediction-based representation learning for general audio, focusing on both task-specific performance and generalization across diverse downstream tasks.
Learned representations were evaluated through linear evaluation and fine-tuning on benchmarks used in M2D\cite{M2D2024TASLP}, and the proposed method was further analyzed via ablation studies on hint ratio scheduling.

Linear evaluation focuses on the linear separability of frozen representations across diverse downstream tasks and also serves as an indicator of generalization.
Fine-tuning evaluates end-to-end performance under task-specific optimization, reflecting practical effectiveness on standard benchmarks.
Our ablation studies examined the sensitivity of DWM to hint ratio scheduling, providing insight into its design behavior.

We evaluated three masking strategies, random masking, IBM, and the proposed DWM, by integrating each into masked prediction-based SSL frameworks.
SGIM was excluded from evaluation due to its significant computational overhead, which renders large-scale pre-training impractical in our experimental setup.

Experiments were conducted using a series of three SSL frameworks: MSM-MAE\cite{niizumi2022msm-mae}, which adapts MAE to audio spectrograms; M2D\cite{M2D2024TASLP}, which learns by predicting masked patch representations; and M2D-CLAP\cite{niizumi2025m2d-clap}, which extends M2D by incorporating semantic supervision from audio-text alignment.
These frameworks share most of the encoder architecture and training pipeline, differing primarily in their learning objectives, which makes them well suited for a controlled comparison in our experiments.

\begin{table*}[htb!]
\caption{Linear evaluation results (\%) with 95\% CI comparing different masking strategies.}
\label{tab:results-le}
\centering
\resizebox{1.0\textwidth}{!}{%
\begin{tabular}{lllllllllll} \toprule
  &  \multicolumn{2}{c}{Env. sound tasks} & \multicolumn{4}{c}{Speech tasks} & \multicolumn{3}{c}{Music tasks} \\
 \cmidrule(lr){2-3} \cmidrule(lr){4-7} \cmidrule(lr){8-10} 
Masking strategy  &   ESC-50 &    US8K &    SPCV2 &    VC1 &     VF &    CRM-D &    GTZAN &     NSynth &      Surge & Avg.\\
\midrule
\multicolumn{10}{l}{\textit{(MSM-MAE variants)}}  \\
Random & 89.2{\fontsize{6pt}{6pt} \selectfont $\pm$0.4} & 87.2{\fontsize{6pt}{6pt} \selectfont $\pm$0.2} &\textbf{96.1{\fontsize{6pt}{6pt} \selectfont $\pm$0.1}}&\textbf{73.4{\fontsize{6pt}{6pt} \selectfont $\pm$0.4}}&\textbf{97.9{\fontsize{6pt}{6pt} \selectfont $\pm$0.1}}&\textbf{71.3{\fontsize{6pt}{6pt} \selectfont $\pm$0.2}}& 80.3{\fontsize{6pt}{6pt} \selectfont $\pm$1.0} & 74.6{\fontsize{6pt}{6pt} \selectfont $\pm$0.4} & 43.2{\fontsize{6pt}{6pt} \selectfont $\pm$0.2} & 79.2\\
\addlinespace[0.5mm]

IBM\cite{Baevski2023data2vec2} &\textbf{90.1{\fontsize{6pt}{6pt} \selectfont $\pm$0.3}}&\textbf{88.2{\fontsize{6pt}{6pt} \selectfont $\pm$0.4}}& 95.4{\fontsize{6pt}{6pt} \selectfont $\pm$0.1} & 65.1{\fontsize{6pt}{6pt} \selectfont $\pm$0.5} & 97.7{\fontsize{6pt}{6pt} \selectfont $\pm$0.1} & 70.9{\fontsize{6pt}{6pt} \selectfont $\pm$0.6} & 80.5{\fontsize{6pt}{6pt} \selectfont $\pm$2.3} & 74.2{\fontsize{6pt}{6pt} \selectfont $\pm$0.6} & 42.8{\fontsize{6pt}{6pt} \selectfont $\pm$0.1} & 78.3\\
\multicolumn{1}{r}{vs. random [pp]} & +0.9 & +1.0 & -0.7 & -8.3 & -0.2 & -0.4 & +0.2 & -0.4 & -0.4 & -0.9 \\
\addlinespace[0.5mm]

DWM (proposed) & 89.9{\fontsize{6pt}{6pt} \selectfont $\pm$0.7} & 87.4{\fontsize{6pt}{6pt} \selectfont $\pm$0.1} & 95.9{\fontsize{6pt}{6pt} \selectfont $\pm$0.1} & 72.5{\fontsize{6pt}{6pt} \selectfont $\pm$0.7} & 97.8{\fontsize{6pt}{6pt} \selectfont $\pm$0.1} & 70.9{\fontsize{6pt}{6pt} \selectfont $\pm$0.3} &\textbf{81.8{\fontsize{6pt}{6pt} \selectfont $\pm$1.1}}&\textbf{75.5{\fontsize{6pt}{6pt} \selectfont $\pm$0.1}}&\textbf{43.3{\fontsize{6pt}{6pt} \selectfont $\pm$0.3}}& \textbf{79.5}\\
\multicolumn{1}{r}{vs. random [pp]} & +0.7 & +0.2 & -0.2 & -0.9 & -0.1 & -0.4 & +1.5 & +0.9 & +0.1 & +0.3 \\

\midrule
\multicolumn{10}{l}{\textit{(M2D variants)}}  \\
Random & 91.3{\fontsize{6pt}{6pt} \selectfont $\pm$0.6} & 87.6{\fontsize{6pt}{6pt} \selectfont $\pm$0.2} & 96.0{\fontsize{6pt}{6pt} \selectfont $\pm$0.1} &\textbf{73.4{\fontsize{6pt}{6pt} \selectfont $\pm$0.2}}& 98.3{\fontsize{6pt}{6pt} \selectfont $\pm$0.0} & 73.0{\fontsize{6pt}{6pt} \selectfont $\pm$0.7} & 84.1{\fontsize{6pt}{6pt} \selectfont $\pm$2.7} &\textbf{75.7{\fontsize{6pt}{6pt} \selectfont $\pm$0.1}}& 42.1{\fontsize{6pt}{6pt} \selectfont $\pm$0.2} & 80.2\\
\addlinespace[0.5mm]

IBM\cite{Baevski2023data2vec2} &\textbf{92.8{\fontsize{6pt}{6pt} \selectfont $\pm$0.3}}&\textbf{88.2{\fontsize{6pt}{6pt} \selectfont $\pm$0.2}}& 95.8{\fontsize{6pt}{6pt} \selectfont $\pm$0.1} & 69.1{\fontsize{6pt}{6pt} \selectfont $\pm$0.3} & 98.3{\fontsize{6pt}{6pt} \selectfont $\pm$0.0} &\textbf{73.2{\fontsize{6pt}{6pt} \selectfont $\pm$0.7}}&\textbf{85.3{\fontsize{6pt}{6pt} \selectfont $\pm$2.0}}& 75.5{\fontsize{6pt}{6pt} \selectfont $\pm$0.6} &\textbf{42.5{\fontsize{6pt}{6pt} \selectfont $\pm$0.2}}& 80.1\\
\multicolumn{1}{r}{vs. random [pp]} & +1.5 & +0.6 & -0.2 & -4.3 & 0.0 & +0.2 & +1.2 & -0.2 & +0.4 & -0.1 \\
\addlinespace[0.5mm]

DWM (proposed) & 91.9{\fontsize{6pt}{6pt} \selectfont $\pm$0.3} & 87.9{\fontsize{6pt}{6pt} \selectfont $\pm$0.2} &\textbf{96.1{\fontsize{6pt}{6pt} \selectfont $\pm$0.1}}& 72.8{\fontsize{6pt}{6pt} \selectfont $\pm$0.4} &\textbf{98.4{\fontsize{6pt}{6pt} \selectfont $\pm$0.0}}& 73.0{\fontsize{6pt}{6pt} \selectfont $\pm$0.8} & 85.1{\fontsize{6pt}{6pt} \selectfont $\pm$1.2} & 75.4{\fontsize{6pt}{6pt} \selectfont $\pm$0.5} & 42.0{\fontsize{6pt}{6pt} \selectfont $\pm$0.3} & \textbf{80.3}\\
\multicolumn{1}{r}{vs. random [pp]} & +0.6 & +0.3 & +0.1 & -0.6 & +0.1 & 0.0 & +1.0 & -0.3 & -0.1 & +0.1 \\

\midrule
\multicolumn{10}{l}{\textit{(M2D-CLAP variants)}}  \\
Random & 95.2{\fontsize{6pt}{6pt} \selectfont $\pm$0.3} & 89.5{\fontsize{6pt}{6pt} \selectfont $\pm$0.2} &\textbf{95.6{\fontsize{6pt}{6pt} \selectfont $\pm$0.1}}&\textbf{68.6{\fontsize{6pt}{6pt} \selectfont $\pm$0.5}}&\textbf{98.1{\fontsize{6pt}{6pt} \selectfont $\pm$0.1}}& 73.4{\fontsize{6pt}{6pt} \selectfont $\pm$0.7} & 86.6{\fontsize{6pt}{6pt} \selectfont $\pm$1.6} &\textbf{76.5{\fontsize{6pt}{6pt} \selectfont $\pm$0.4}}& 42.2{\fontsize{6pt}{6pt} \selectfont $\pm$0.3} & 80.6\\
\addlinespace[0.5mm]

IBM\cite{Baevski2023data2vec2} &\textbf{95.8{\fontsize{6pt}{6pt} \selectfont $\pm$0.2}}& 89.5{\fontsize{6pt}{6pt} \selectfont $\pm$0.2} & 95.2{\fontsize{6pt}{6pt} \selectfont $\pm$0.1} & 65.7{\fontsize{6pt}{6pt} \selectfont $\pm$0.2} & 97.9{\fontsize{6pt}{6pt} \selectfont $\pm$0.1} &\textbf{73.6{\fontsize{6pt}{6pt} \selectfont $\pm$0.6}}& 86.4{\fontsize{6pt}{6pt} \selectfont $\pm$0.9} & 76.5{\fontsize{6pt}{6pt} \selectfont $\pm$0.6} & 42.5{\fontsize{6pt}{6pt} \selectfont $\pm$0.2} & 80.3\\
\multicolumn{1}{r}{vs. random [pp]} & +0.6 & 0.0 & -0.4 & -2.9 & -0.2 & +0.2 & -0.2 & 0.0 & +0.3 & -0.3 \\
\addlinespace[0.5mm]

DWM (proposed) & 95.6{\fontsize{6pt}{6pt} \selectfont $\pm$1.0} &\textbf{90.2{\fontsize{6pt}{6pt} \selectfont $\pm$0.4}}& 95.5{\fontsize{6pt}{6pt} \selectfont $\pm$0.1} & 67.9{\fontsize{6pt}{6pt} \selectfont $\pm$0.5} & 98.0{\fontsize{6pt}{6pt} \selectfont $\pm$0.0} & 73.2{\fontsize{6pt}{6pt} \selectfont $\pm$1.8} &\textbf{87.7{\fontsize{6pt}{6pt} \selectfont $\pm$4.3}}& 75.2{\fontsize{6pt}{6pt} \selectfont $\pm$0.4} &\textbf{43.0{\fontsize{6pt}{6pt} \selectfont $\pm$0.2}}& \textbf{80.7}\\
\multicolumn{1}{r}{vs. random [pp]} & +0.4 & +0.7 & -0.1 & -0.7 & -0.1 & -0.2 & +1.1 & -1.3 & +0.8 & +0.1 \\

\midrule
\multicolumn{10}{l}{\textit{(Reference: Results of prior methods evaluated in this work)}}  \\
BEATs\cite{chen2022beats} (Random) & 86.9{\fontsize{6pt}{6pt} \selectfont $\pm$1.4} & 84.8{\fontsize{6pt}{6pt} \selectfont $\pm$0.1} & 89.4{\fontsize{6pt}{6pt} \selectfont $\pm$0.1} & 41.4{\fontsize{6pt}{6pt} \selectfont $\pm$0.7} & 94.1{\fontsize{6pt}{6pt} \selectfont $\pm$0.3} & 64.7{\fontsize{6pt}{6pt} \selectfont $\pm$0.8} & 72.6{\fontsize{6pt}{6pt} \selectfont $\pm$4.3} & 75.9{\fontsize{6pt}{6pt} \selectfont $\pm$0.2} & 39.3{\fontsize{6pt}{6pt} \selectfont $\pm$0.4} & 72.1 \\

EAT\cite{EAT_SSL} (IBM) & 85.6{\fontsize{6pt}{6pt} \selectfont $\pm$0.4} & 81.7{\fontsize{6pt}{6pt} \selectfont $\pm$0.3} & 81.5{\fontsize{6pt}{6pt} \selectfont $\pm$0.4} & 39.6{\fontsize{6pt}{6pt} \selectfont $\pm$0.5} & 92.6{\fontsize{6pt}{6pt} \selectfont $\pm$0.0} & 64.9{\fontsize{6pt}{6pt} \selectfont $\pm$2.8} & 73.7{\fontsize{6pt}{6pt} \selectfont $\pm$0.5} & 71.9{\fontsize{6pt}{6pt} \selectfont $\pm$0.1} & 39.0{\fontsize{6pt}{6pt} \selectfont $\pm$0.3} & 70.0 \\

\bottomrule
\end{tabular}
}
\vspace{-10pt}
\end{table*}

\subsection{Experimental Setup}
All pre-training settings followed those in the base SSL methods, including a masking ratio of 0.75 for MSM-MAE and 0.7 for M2D and M2D-CLAP.
We pre-trained two models per masking strategy for MSM-MAE and M2D, while only one model per strategy was trained for M2D-CLAP due to resource constraints.
For DWM, the hint scheduling parameter $\gamma$ was set to 2, coinciding with the value reported in SemMAE\cite{Li2022SemMAE}. The hint ratio was initialized at 1.0 and gradually annealed to zero, as illustrated in Fig.~\ref{fig:hint_ratio}.
For IBM, we followed EAT\cite{EAT_SSL} for the block size (i.e., $5 \times 5$).

All evaluation settings followed those in \cite{niizumi2023byol-a,niizumi2022msm-mae,M2D2024TASLP,niizumi2025m2d-clap}, including the evaluation platform (EVAR\footnote{\url{https://github.com/nttcslab/eval-audio-repr}}) and the downstream classification tasks spanning environmental sound, speech, and music.
For each model and downstream task, we conducted three runs and report the resulting statistics as the final results. When two models are pre-trained, this results in six evaluations per pre-training setting.

Environmental sound tasks include AudioSet\cite{gemmeke2017audioset}, ESC-50\cite{piczak2015esc50}, UrbanSound8K\cite{salamon2014urbansound} (US8K); speech tasks include Speech Commands V2\cite{speechcommandsv2} (SPCV2), VoxCeleb1\cite{voxceleb} (VC1), VoxForge\cite{voxforge} (VF), and CREMA-D\cite{cao2014cremad} (CRM-D); and music tasks include GTZAN\cite{gt2013gtzan}, NSynth\cite{nsynth2017}, and the Pitch Audio Dataset\cite{turian2021torchsynth} (Surge).
AudioSet is evaluated under two settings: AS2M using the full 2M training samples and AS20K using 21K samples from the balanced training set. Surge is a pitch classification task over 88 MIDI notes.

All tasks are formulated as classification problems, and performance is reported in terms of accuracy or mean average precision (mAP).

\begin{table}[tb!]
\caption{Fine-tuning results with 95\% CI comparing masking strategies.}
\label{tab:results-ft}
\centering
\resizebox{1.0\columnwidth}{!}{%
\begin{tabular}{llllll} \toprule
Masking  & AS2M & AS20K &     ESC-50 &  SPCV2 &       VC1\\
\vspace{-1pt} strategy  & mAP & mAP &  acc(\%) &  acc(\%) &    acc(\%)\\
\midrule
\multicolumn{6}{l}{\textit{(MAE variants)}}  \\
\addlinespace[0.05cm]
Random & 47.4{\fontsize{6pt}{6pt} \selectfont $\pm$0.1} & 37.9{\fontsize{6pt}{6pt} \selectfont $\pm$0.0} & 95.4{\fontsize{6pt}{6pt} \selectfont $\pm$0.1} &\textbf{98.4{\fontsize{6pt}{6pt} \selectfont $\pm$0.0}}&\textbf{96.6{\fontsize{6pt}{6pt} \selectfont $\pm$0.2}}\\
\addlinespace[0.5mm]

IBM\cite{Baevski2023data2vec2} & 47.4{\fontsize{6pt}{6pt} \selectfont $\pm$0.1} &\textbf{38.6{\fontsize{6pt}{6pt} \selectfont $\pm$0.1}}& 95.4{\fontsize{6pt}{6pt} \selectfont $\pm$0.3} & 98.3{\fontsize{6pt}{6pt} \selectfont $\pm$0.0} & 95.4{\fontsize{6pt}{6pt} \selectfont $\pm$0.5} \\
\multicolumn{1}{r}{diff.[pp]} & 0.0 & +0.7 & 0.0 & -0.1 & -1.2 \\
\addlinespace[0.5mm]

DWM &\textbf{47.5{\fontsize{6pt}{6pt} \selectfont $\pm$0.1}}& 38.3{\fontsize{6pt}{6pt} \selectfont $\pm$0.1} &\textbf{95.5{\fontsize{6pt}{6pt} \selectfont $\pm$0.2}}& \textbf{98.4{\fontsize{6pt}{6pt} \selectfont $\pm$0.1}} & 96.4{\fontsize{6pt}{6pt} \selectfont $\pm$0.2} \\
\multicolumn{1}{r}{diff.[pp]} & +0.1 & +0.4 & +0.1 & 0.0 & -0.2 \\

\midrule
\multicolumn{6}{l}{\textit{(M2D variants)}}  \\
\addlinespace[0.05cm]
Random & 47.8{\fontsize{6pt}{6pt} \selectfont $\pm$0.1} & 38.6{\fontsize{6pt}{6pt} \selectfont $\pm$0.1} & 96.0{\fontsize{6pt}{6pt} \selectfont $\pm$0.2} &\textbf{98.4{\fontsize{6pt}{6pt} \selectfont $\pm$0.1}}&\textbf{96.3{\fontsize{6pt}{6pt} \selectfont $\pm$0.2}}\\
\addlinespace[0.5mm]

IBM\cite{Baevski2023data2vec2} &\textbf{47.9{\fontsize{6pt}{6pt} \selectfont $\pm$0.2}}&\textbf{39.3{\fontsize{6pt}{6pt} \selectfont $\pm$0.1}}&\textbf{96.2{\fontsize{6pt}{6pt} \selectfont $\pm$0.2}}& \textbf{98.4{\fontsize{6pt}{6pt} \selectfont $\pm$0.0}} & 95.4{\fontsize{6pt}{6pt} \selectfont $\pm$0.7} \\
\multicolumn{1}{r}{diff.[pp]} & +0.1 & +0.7 & +0.2 & 0.0 & -0.9 \\
\addlinespace[0.5mm]

DWM & 47.8{\fontsize{6pt}{6pt} \selectfont $\pm$0.0} & 38.9{\fontsize{6pt}{6pt} \selectfont $\pm$0.0} & 96.0{\fontsize{6pt}{6pt} \selectfont $\pm$0.2} & \textbf{98.4{\fontsize{6pt}{6pt} \selectfont $\pm$0.1}} & 96.1{\fontsize{6pt}{6pt} \selectfont $\pm$0.2} \\
\multicolumn{1}{r}{diff.[pp]} & 0.0 & +0.3 & 0.0 & 0.0 & -0.2 \\

\midrule
\multicolumn{6}{l}{\textit{(M2D-CLAP variants)}}  \\
\addlinespace[0.05cm]
Random & 49.0{\fontsize{6pt}{6pt} \selectfont $\pm$0.1} & 42.1{\fontsize{6pt}{6pt} \selectfont $\pm$0.0} & 97.8{\fontsize{6pt}{6pt} \selectfont $\pm$0.1} & 98.4{\fontsize{6pt}{6pt} \selectfont $\pm$0.0} &\textbf{95.5{\fontsize{6pt}{6pt} \selectfont $\pm$0.3}}\\
\addlinespace[0.5mm]

IBM\cite{Baevski2023data2vec2} &\textbf{49.1{\fontsize{6pt}{6pt} \selectfont $\pm$0.1}}&\textbf{42.3{\fontsize{6pt}{6pt} \selectfont $\pm$0.1}}&\textbf{98.0{\fontsize{6pt}{6pt} \selectfont $\pm$0.1}}& 98.4{\fontsize{6pt}{6pt} \selectfont $\pm$0.1} & 94.2{\fontsize{6pt}{6pt} \selectfont $\pm$0.1} \\
\multicolumn{1}{r}{diff.[pp]} & +0.1 & +0.2 & +0.2 & 0.0 & -1.3 \\
\addlinespace[0.5mm]

DWM & 49.0{\fontsize{6pt}{6pt} \selectfont $\pm$0.2} & 41.8{\fontsize{6pt}{6pt} \selectfont $\pm$0.1} & 97.8{\fontsize{6pt}{6pt} \selectfont $\pm$0.1} &\textbf{98.5{\fontsize{6pt}{6pt} \selectfont $\pm$0.0}}& 95.1{\fontsize{6pt}{6pt} \selectfont $\pm$0.4} \\
\multicolumn{1}{r}{diff.[pp]} & 0.0 & -0.3 & 0.0 & +0.1 & -0.4 \\

\midrule
\multicolumn{6}{l}{\textit{(Reference: Performance reported in prior studies)}}  \\
\addlinespace[0.05cm]
BEATs\cite{chen2022beats} & 48.0 & 38.3 & 95.6 & 98.3 & - \\
EAT\cite{EAT_SSL} & 48.6 & 40.2 & 95.9 & 98.3 & - \\

\bottomrule
\end{tabular}
}
\vspace{-10pt}
\end{table}

\begin{table*}[tb!]
\caption{Ablation of hint ratio scheduling in DWM: Linear evaluation results (\%) with 95\% CI.}
\label{tab:results-abl}
\centering
\resizebox{1.0\textwidth}{!}{%
\begin{tabular}{lllllllllll} \toprule
  &  \multicolumn{2}{c}{Env. sound tasks} & \multicolumn{4}{c}{Speech tasks} & \multicolumn{3}{c}{Music tasks} \\
 \cmidrule(lr){2-3} \cmidrule(lr){4-7} \cmidrule(lr){8-10} 
Scheduling  &   ESC-50 &    US8K &    SPCV2 &    VC1 &     VF &    CRM-D &    GTZAN &     NSynth &      Surge & Avg.\\
\midrule

Baseline MSM-MAE & 89.2{\fontsize{6pt}{6pt} \selectfont $\pm$0.4} & 87.2{\fontsize{6pt}{6pt} \selectfont $\pm$0.2} & 96.1{\fontsize{6pt}{6pt} \selectfont $\pm$0.1} & \textbf{73.4{\fontsize{6pt}{6pt} \selectfont $\pm$0.4}} & 97.9{\fontsize{6pt}{6pt} \selectfont $\pm$0.1} & 71.3{\fontsize{6pt}{6pt} \selectfont $\pm$0.2} & 80.3{\fontsize{6pt}{6pt} \selectfont $\pm$1.0} & 74.6{\fontsize{6pt}{6pt} \selectfont $\pm$0.4} & 43.2{\fontsize{6pt}{6pt} \selectfont $\pm$0.2} & 79.2\\
\midrule
Hint ratio $r_h$=0.0 &\textbf{89.9{\fontsize{6pt}{6pt} \selectfont $\pm$1.1}}& 87.3{\fontsize{6pt}{6pt} \selectfont $\pm$0.7} & 95.5{\fontsize{6pt}{6pt} \selectfont $\pm$0.1} & 69.1{\fontsize{6pt}{6pt} \selectfont $\pm$0.1} & 97.7{\fontsize{6pt}{6pt} \selectfont $\pm$0.0} & 70.4{\fontsize{6pt}{6pt} \selectfont $\pm$0.6} &\textbf{83.4{\fontsize{6pt}{6pt} \selectfont $\pm$4.5}}& 75.4{\fontsize{6pt}{6pt} \selectfont $\pm$0.2} & 42.5{\fontsize{6pt}{6pt} \selectfont $\pm$0.9} & 79.0\\
Hint ratio $r_h$=0.5 & 87.4{\fontsize{6pt}{6pt} \selectfont $\pm$0.9} & 86.3{\fontsize{6pt}{6pt} \selectfont $\pm$0.6} & 96.0{\fontsize{6pt}{6pt} \selectfont $\pm$0.2} & 71.8{\fontsize{6pt}{6pt} \selectfont $\pm$0.5} & 97.6{\fontsize{6pt}{6pt} \selectfont $\pm$0.3} & 70.5{\fontsize{6pt}{6pt} \selectfont $\pm$0.9} & 82.1{\fontsize{6pt}{6pt} \selectfont $\pm$3.1} & 74.3{\fontsize{6pt}{6pt} \selectfont $\pm$0.4} & 42.5{\fontsize{6pt}{6pt} \selectfont $\pm$0.5} & 78.7\\
Schedule $\gamma$=1 & 88.1{\fontsize{6pt}{6pt} \selectfont $\pm$0.2} & 87.3{\fontsize{6pt}{6pt} \selectfont $\pm$0.3} &\textbf{96.2{\fontsize{6pt}{6pt} \selectfont $\pm$0.2}}& 70.0{\fontsize{6pt}{6pt} \selectfont $\pm$0.5} & 97.7{\fontsize{6pt}{6pt} \selectfont $\pm$0.2} &\textbf{71.4{\fontsize{6pt}{6pt} \selectfont $\pm$0.9}}& 82.0{\fontsize{6pt}{6pt} \selectfont $\pm$6.4} &\textbf{76.0{\fontsize{6pt}{6pt} \selectfont $\pm$0.2}}& 42.6{\fontsize{6pt}{6pt} \selectfont $\pm$0.7} & 79.0\\
Schedule $\gamma$=2 & \textbf{89.9{\fontsize{6pt}{6pt} \selectfont $\pm$0.7}} &\textbf{87.4{\fontsize{6pt}{6pt} \selectfont $\pm$0.1}}& 95.9{\fontsize{6pt}{6pt} \selectfont $\pm$0.1} & 72.5{\fontsize{6pt}{6pt} \selectfont $\pm$0.7} & 97.8{\fontsize{6pt}{6pt} \selectfont $\pm$0.1} & 70.9{\fontsize{6pt}{6pt} \selectfont $\pm$0.3} & 81.8{\fontsize{6pt}{6pt} \selectfont $\pm$1.1} & 75.5{\fontsize{6pt}{6pt} \selectfont $\pm$0.1} &\textbf{43.3{\fontsize{6pt}{6pt} \selectfont $\pm$0.3}}& \textbf{79.5}\\
Schedule $\gamma$=4 & 89.5{\fontsize{6pt}{6pt} \selectfont $\pm$0.8} & 87.3{\fontsize{6pt}{6pt} \selectfont $\pm$0.3} & 95.9{\fontsize{6pt}{6pt} \selectfont $\pm$0.2} &\textbf{73.4{\fontsize{6pt}{6pt} \selectfont $\pm$0.5}}& \textbf{98.0{\fontsize{6pt}{6pt} \selectfont $\pm$0.1}} & 69.9{\fontsize{6pt}{6pt} \selectfont $\pm$1.9} & 81.7{\fontsize{6pt}{6pt} \selectfont $\pm$4.5} & 74.7{\fontsize{6pt}{6pt} \selectfont $\pm$0.2} & 43.1{\fontsize{6pt}{6pt} \selectfont $\pm$0.4} & 79.3\\
Schedule $\gamma$=8 & 89.5{\fontsize{6pt}{6pt} \selectfont $\pm$2.7} & 87.2{\fontsize{6pt}{6pt} \selectfont $\pm$0.6} & \textbf{96.2{\fontsize{6pt}{6pt} \selectfont $\pm$0.0}} & \textbf{73.4{\fontsize{6pt}{6pt} \selectfont $\pm$0.3}} &\textbf{98.0{\fontsize{6pt}{6pt} \selectfont $\pm$0.2}}& 70.1{\fontsize{6pt}{6pt} \selectfont $\pm$2.0} & 82.5{\fontsize{6pt}{6pt} \selectfont $\pm$5.8} & 72.5{\fontsize{6pt}{6pt} \selectfont $\pm$0.7} & 42.9{\fontsize{6pt}{6pt} \selectfont $\pm$0.4} & 79.2\\

\bottomrule
\end{tabular}
}
\vspace{-10pt}
\end{table*}

\subsection{Comparison of Masking Strategies in Linear Evaluation}
With the linear evaluation setup, we compared masking strategies by assessing the linear separability of frozen representations across downstream tasks, which also serves as an indicator of representation generalization.

The results in Table~\ref{tab:results-le} indicate that both IBM and DWM influence downstream task performance consistently across different base SSL frameworks.
Their effects vary by task domain: performance improvements are generally observed for environmental sound tasks, whereas degradations tend to appear in speech tasks. For music tasks, the effect differs across tasks: performance improves for genre classification (GTZAN) but degrades for instrument classification (NSynth). Although the variance across tasks is non-negligible, IBM overall exhibits a larger impact on performance compared to DWM.

IBM particularly improves performance on environmental sound tasks (ESC-50 and US8K) by approximately 1 pp, which is consistent with the performance gains on AudioSet reported in EAT that employs IBM. In contrast, notable performance degradation is observed for speaker identification on VC1, with drops of up to 8.3 pp in MSM-MAE and 4.3 pp in M2D. This suggests that IBM may suppress fine-grained speaker-specific cues that are crucial for tasks requiring sensitivity to subtle vocal differences.

DWM exhibits relatively smaller gains on ESC-50 and US8K compared to IBM, while being characterized by notably smaller performance degradation on speech tasks.
Although the improvements are modest, DWM consistently improves performance on ESC-50, US8K, and GTZAN, contributing to overall performance gains across tasks.
Performance degradation on speech tasks is largely mitigated, with the drop on VC1 remaining within 1 pp, which is substantially smaller than that observed with IBM.
Overall, these results suggest that DWM modestly improves performance while better preserving generalization across diverse downstream tasks.

\subsection{Comparison of Masking Strategies in Fine-tuning}
In fine-tuning comparisons shown in Table~\ref{tab:results-ft}, the impact of different masking strategies is generally smaller than that observed in linear evaluation; nevertheless, similar trends can still be identified, with performance improvements on environmental sound tasks and notable degradation on VC1.

IBM yields little gain on AS2M but more consistent gains on AS20K, where training data are limited, suggesting that it facilitates learning representations that generalize across diverse patterns of audio events.
Conversely, the substantial performance drop on speaker identification (VC1) indicates a trade-off with the fine-grained discriminative information required for this task.

Compared to IBM, DWM exhibits more moderate performance changes across tasks, while consistently reducing degradation on VC1, in line with the observations from linear evaluation.
On AS20K, DWM generally maintains favorable performance for MSM-MAE and M2D, indicating improved performance with preserved generalization.
In contrast, AS20K performance decreases for M2D-CLAP; although based on limited runs due to computational constraints, it suggests that the effectiveness of DWM may be sensitive to training conditions when competing for high-end performance.

\begin{figure}[tbp]
  \centering
  \includegraphics[width=1.0\columnwidth]{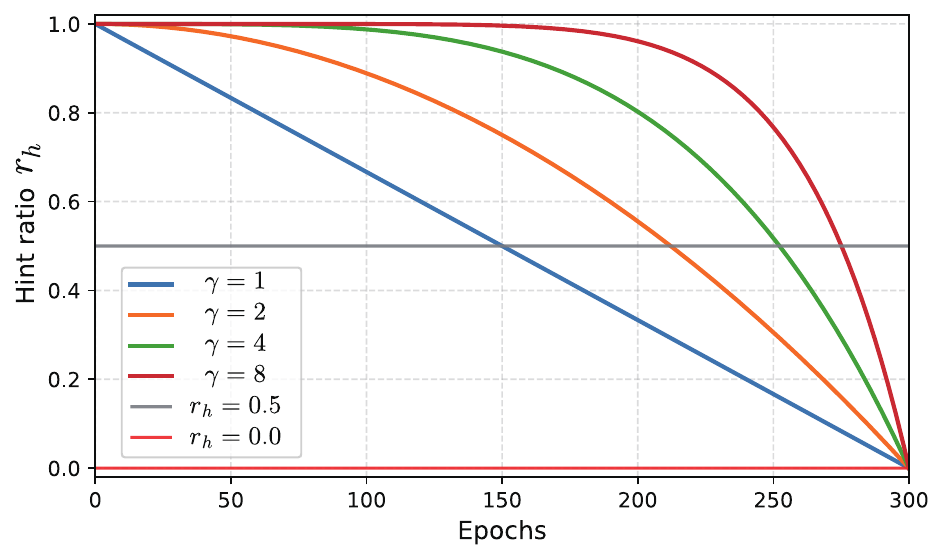}
  \vspace{-20pt}
  \caption{Hint ratio $r_h$ schedules over training epochs for ablation settings.}
  \label{fig:hint_ratio}
  \vspace{-10pt}
\end{figure}

\subsection{Ablations on DWM Hint Ratio Scheduling}
We conducted an ablation study on the scheduling of the hint ratio $r_h$ in DWM, which is designed to prevent the prediction task from becoming overly difficult in the early stages of training.
As illustrated in Fig.~\ref{fig:hint_ratio}, we either fixed $r_h$ to a constant value or varied the scheduling parameter $\gamma$, and evaluated the resulting MSM-MAE models using linear evaluation.
The results are summarized in Table~\ref{tab:results-abl}.

When $r_h$ was fixed to 0.0, some tasks showed improved performance; however, a substantial degradation was observed on VC1.
When $r_h$ was fixed to 0.5, performance improvements were observed only on GTZAN, while degradation was evident on the remaining tasks.

In contrast, scheduling $r_h$ using different values of $\gamma$ resulted in more stable behavior.
From $\gamma=2$, degradation on VC1 was effectively mitigated, and with $\gamma=8$, performance became nearly comparable to that of random masking.
Consistent with observations from SG-MAE\cite{Jeongwoo2024SelfGuided} and SemMAE\cite{Li2022SemMAE}, these results highlight the importance of appropriately scheduling task difficulty during training, which also holds for DWM.

\subsection{Summary of Experiments}
The experiments reveal several insights into the role of masking strategies.
First, masking strategies substantially influence downstream performance across tasks, even when the underlying SSL framework is fixed, indicating that masking design is a non-trivial factor in masked prediction-based representation learning.
Second, while IBM can yield notable gains on environmental sound tasks, it tends to incur a trade-off with fine-grained discriminative tasks such as speaker identification, leading to degraded generalization.
In contrast, DWM achieves more moderate yet consistent improvements across tasks, while showing less performance degradation on speech-related tasks.
Finally, the ablation results highlight the importance of appropriately scheduling task difficulty via the hint ratio, demonstrating that controlled masking progression is critical for stable and generalizable representation learning.

\section{Conclusion}
In this work, we investigated the role of masking strategies in masked prediction-based self-supervised learning for general audio using spectrogram inputs.
Through comparisons across multiple SSL frameworks and downstream tasks, we showed that masking strategies substantially influence both task performance and generalization. 

Our analysis revealed that IBM substantially improves environmental sound performance under a limited-data condition, while reducing generalization, particularly to speech tasks.
In our experiments, informed masking approaches such as SGIM exhibited non-negligible computational overhead.
Motivated by these observations, we proposed DWM, a lightweight alternative that exploits dispersion characteristics inherent to audio spectrograms.

Experimental results demonstrated that DWM yields modest but consistent performance improvements while largely maintaining performance on speech tasks, thereby better preserving generalization across diverse downstream tasks.
In addition, ablation studies highlighted the importance of scheduling the hint ratio to control task difficulty during training for stable and effective representation learning.

Overall, our findings provide empirical insights into masking strategy design and suggest that lightweight, spectrogram-aware masking can serve as a practical alternative to computationally intensive informed masking approaches. Future work includes exploring dispersion metrics beyond MAD that better capture semantically relevant content, evaluating DWM with varying patch sizes and larger-scale datasets, and extending comparisons to computationally intensive methods such as SGIM under matched budgets.

\bibliographystyle{IEEEtran}
\bibliography{refs}

\appendices

\section{Notes on the Design of Dispersion-Weighted Masking}
\label{app:theory}

Audio events typically excite specific frequency bands, producing patches with high spectral variability in active regions and low variability in silent or tonal regions. MAD directly captures this variability: a high MAD value indicates that the patch contains a mixture of active and inactive frequency components, which correlates with the presence of informative spectral content. This property makes MAD a computationally efficient proxy for patch-level information density without requiring explicit object detection or eigenvalue decomposition.

We attribute the balanced cross-task behavior of DWM to its probabilistic nature: unlike IBM, which deterministically unmasks contiguous blocks, DWM distributes masking probability across all patches proportionally to their dispersion. This preserves a degree of randomness that maintains sensitivity to fine-grained features (e.g., speaker characteristics in VC1), while still biasing the masking toward spectrally informative regions that benefit event-related tasks.

\section{Limitations of MAD as a Dispersion Metric}
\label{app:limitations}

One limitation of using MAD as a dispersion metric is that it measures within-patch variability rather than absolute energy or semantic relevance. Consequently, a patch containing a strong but spectrally uniform signal (e.g., a sustained pure tone) may receive a low dispersion score despite carrying meaningful content. Conversely, patches with high variability due to noise may be overweighted. Investigating alternative or complementary importance metrics remains an avenue for future work.

\end{document}